\documentclass[12pt]{article}
\usepackage{amsmath}
\usepackage{cite}
\usepackage{relsize}
\usepackage{amssymb}
\usepackage{graphics}
\usepackage{epsfig}
\usepackage{epstopdf}
\usepackage{verbatim}
\usepackage{color}
\usepackage{xcolor}
\usepackage{subcaption}
\usepackage{multirow}
\usepackage{textcomp}
\usepackage{eqnarray}
\usepackage{float}
\usepackage{mathtools}
\usepackage{pifont}
\newcommand{\cmark}{\ding{51}}%
\newcommand{\xmark}{\ding{55}}%
\usepackage[toc,page]{appendix}
\usepackage{enumitem}
\usepackage{ragged2e}
\usepackage{dirtytalk}
\usepackage{mathrsfs}
\renewcommand{\today}{}
 \font\tenrm=cmr12
\begin{document}

\title{\textbf{Leptogenesis and Neutrinoless Double Beta in the Scotogenic
  Hybrid Textures of Neutrino Mass Matrix}}

 \author {Ankush\thanks{ ankush.bbau@gmail.com}, Rishu Verma\thanks{ rishuvrm274@gmail.com }, Sahil Kumar\thanks{ sahilruhaan1999@gmail.com }   and B. C. Chauhan\thanks{ bcawake@hpcu.ac.in}}

\date{\textit{Department of Physics and Astronomical Science,\\Central University of Himachal Pradesh, Dharamshala 176215, INDIA.}}

\maketitle
\begin{abstract}
\noindent In our recent work we identify the hybrid textures which simultaneously account for dark matter (DM), neutrinoless double beta decay ($0\nu\beta\beta$) and leptogenesis. We also obtained the bounds on dark matter mass and effective Majorana mass $|M_{ee}|$.  We have found correlation of baryon asymmetry of universe $Y$ with dark matter mass $M_1$ and effective Majorana mass $|M_{ee}|$. We use experimental bounds on relic density of dark matter ($\Omega h^2$) and baryon asymmetry of universe to identify the hybrid textures. We found that out of five hybrid textures which simultaneously satisfies the physics observations of the DM and $0\nu\beta\beta$ only three hybrid textures altogather satisfy the DM, $0\nu\beta\beta$ and leptogenesis. It is interesting to note that these three hybrid textures gives lower bound to the effective Majorana mass $|M_{ee}|$ which can be probed in current and future experiments like SuperNEMO, KamLAND-Zen, NEXT, and nEXO (5 year) have sensitivity reaches of 0.05 eV, 0.045 eV, 0.03 eV, and 0.015 eV, respectively.
      
\end{abstract}
\section{Introduction}
We  are aware of recent developments in the theory and experiments  of fundamental particles and their interactions at electroweak scale. In the present scenario, it is unquestionably incomplete to solve the puzzels like non vanishing neutrino mass, existence of dark matter  and  baryon asymmetry of universe (BAU). Therefore, a conclusive theory to accommodate all these phenomena  hints towards the physics beyond  standard model (BSM). The history of neutrino oscillation experiments over the last few decades has confirmed that neutrinos are not massless but have a tiny mass. However, these oscillation experiments are only sensitive to the mass square differences and inert to absolute mass scale of the neutrinos. Moreover, the upper limit to the sum of neutrino mass is $\sum M_\nu \leq 0.12$ eV \cite{Planck:2018vyg}.

\noindent The data from cosmological and astrophysical observations of gravitational lensing, glaxy rotation curves, cosmic microwave background (CMBR) and large scale structure formation etc confirm the evidences of the dark matter in the universe. With current abundance 
\begin{equation*}
    \Omega h^2 =0.120 \pm 0.001,
\end{equation*} DM contributs to the $26.8\%$  of the total energy density of the present universe as evident from PLANCK \cite{Planck:2018vyg}. As we know, we don't have any formalism in  SM  to include DM so that it would be stable on the cosmological time scale. However, there exist some BSM frameworks which could stabilize the DM by the inclusion of additional symmetries to the radiative or tree-level models \cite{Verma:2023ahz,Verma:2021koo}. One of such models is at one loop level which we are going to  implement in our study is scotogenic model that simultaneously explains the DM and small neutrino mass \cite{Ma1}.  

\noindent There is ambiguity that whether neutrinos are four component Dirac particles or two component Majorana particles. This ambiguity will become clear once experiments confirm or reject neutrinoless double beta ($0\nu\beta\beta$) decay. This decay would also confirm that the neutrino is its own antiparticle. Furthermore, it would be evidence of lepton number violation and could aid in determining the absolute neutrino mass scale. In fact, there is currently no experimental evidence for this decay. However, with the advancement of future experimental techniques, we hope to see some interesting results.

\noindent The Big Bang theory proposes symmetric initial conditions for particles and antiparticles.
Therefore, as a result of the annihilation effects, no matter would remain. However, what we observe in today's universe is matter dominance, which is referred to as baryon asymmetry of the universe.
Non-symmetric initial conditions, on the other hand, are ineffective in the Standard Model of elementary particle physics due to the inflationary phase and nonperturbative effects. Hence, it is clear that this asymmetry is dynamically generated. However, we don't know where it originated from. The current cosmological bound for baryon asymmetry is \cite{Planck:2018vyg}
\begin{equation*}
    Y=(6.04\pm 0.08)\times10^{-10}.
\end{equation*}%One of the famous formalisms for the generation of lepton asymmetry is the BSM decay of right-handed neutrinos into lepton and Higgs. 

\noindent The small neutrino mass can be generated primarily in two ways, 1) by using a traditional tree-level seesaw models \cite{Minkowski:1977sc,Mohapatra:1979ia,TY,Glashow:1979nm,Gell11}, and 2) by using a radiative mechanism. The radiative models are more promising because the new particles introduced in these models are very much  lighter in comparison to  particles involved in the conventional seesaw mechanism. Radiative neutrino mass generation is possible at one-loop \cite{Ma2,Zee:1980ai,Ma:2001mr,Kubo:2006yx,Hambye:2006zn,Farzan:2009ji}, two-loop \cite{Ma:2007yx,Ma:2007gq,Kajiyama:2013zla,Aoki:2013gzs,Kajiyama:2013rla}, three-loop \cite{Krauss:2002px,Aoki:2008av,Gustafsson:2012vj,Ahriche:2014cda,Ahriche:2014oda,Nomura:2016seu}, and higher-loop levels. The interesting feature of radiative model is the suppression in the mass of the neutrino at n-loop level varies as $\text{(suppression)}^{\text{n}}$ \cite{Bonnet:2012kz}. As a result, with larger loops, these particles can be easily tested at the LHC. In our investigation, we employ the well-known  radiative model called scotogenic model for neutrino mass generation \cite{Ma1}.  The merit of using this model is that it simultaneously account for the  neutrino mass and dark matter. The  implications of the model to the texture zeros are discussed in Refs \cite{Kitabayashi:2018bye,Kitabayashi:2017sjz,Ankush:2021opd}.

\noindent In this work, we make an effort to explain the dark matter, neutrinoless double beta decay and leptogenesis in a single framework under the assumption that the neutrino is a Majorana particle. We operate within a more restrictive framework in the one loop level and hybrid textures in the neutrino mass matrix. Hybrid textures are the ones with  one `zero' and one `equality' in the neutrino mass matrix \cite{Kaneko,Dev:2009he,Goswami:2008uv,Liu:2013oxa,Kalita:2015tda}. Here, all elements of neutrino mass matrix are expressed as proportional to the  non-zero $\left|M_{ee}\right|$ element of the neutrino mass matrix as discussed in our previous work \cite{Ankush:2021opd}. This is one of the key characteristics of hybrid textures taken into consideration in this work.

\noindent The paper is organized as follows: In Section 2, we outline the Scotogenic model and relic density of the dark
matter, The  Section 3 discusses the  dark matter and $0\nu\beta\beta$ decay in the  hybrid textures of neutrino mass matrix and the lepton asymmetry is discussed in section 4. Section 5 discusses the specifics of the numerical analysis and interpretations of  the phenomenological results. Finally, in Section 6, the conclusions are explained.

\section{Scotogenic model and relic density of the dark matter }

In this model, we add three right-handed singlet fermions  $N_{k}  (k=1,2,3)$ which are singlet under $SU(2)_L$ and an $SU(2)_{L}$ scalar doublet $(\eta^{+},\eta^{0})$ to the standard model (SM) which are odd under exact $Z_2$ symmetry  \cite{Ma1}. The  particle content of the  model under $SU(2)_{L}\times U(1)_{Y} \times Z_{2}$ is given by 
\begin{equation}
\begin{split}
    L_{\alpha}&=(\nu_{\alpha},l_{\alpha}):(2,-1/2,+), l^{C}_{\alpha}:(1,1,+),\\
\phi&=(\phi^{+},\phi^{0}):(2,-1/2,+),\\
\eta&=(\eta^{+},\eta^{0}):(2,1/2,-),N_{k}:(1,0,-),
\end{split}
\end{equation}
where $\alpha=e,\mu,\tau$,  $(\nu_{\alpha},l_{\alpha})$ and $(\phi^{+},\phi^{0})$ are left-handed lepton doublets and Higgs doublet, respectively.\\
The Lagrangian of the  model with relevant Yukawa and mass terms is given by
\begin{equation}
    \mathcal{L} \supset Y_{\alpha k}(\bar{\nu}_{\alpha L}\eta^{0}-\bar{l}_{\alpha L} \eta^{+})N_{k} +\frac{1}{2}M_{k}\bar{N}_{k}N^{C}_{k} +H.c.,
\end{equation}
and the relevant scalar potential interaction terms are given by
\begin{equation}
    V\supset \frac{1}{2}\lambda (\phi^{\dagger}\eta)^{2} +h.c.,
\end{equation}
where $\lambda$ is the quartic coupling.
The presence of exact $Z_{2}$ symmetry forbids the neutrinos to acquire mass at tree level and neutrino mass is thus  generated readiatively at one loop level. The general element of neutrino mass matrix $M_{\alpha \beta}$ is given by 
\begin{equation}
\label{I}
    M_{\alpha \beta} = \sum _{k=1}^{3} Y_{\alpha k}Y_{\beta k}\Lambda_{k},
\end{equation}
where
\begin{equation}
    \Lambda_{k}=\frac{\lambda v^{2}}{16 \pi^{2}}\frac{M_{k}}{m_{0}^{2}-M_{k}^{2}}\left(1-\frac{M_{k}^{2}}{m_{0}^{2}-M_{k}^{0}}\ln\frac{m_{0}^{2}}{M_{k}^{2}}\right),
\end{equation}
\begin{equation}
    m_{0}^{2}=\frac{1}{2}(m_{R}^{2}+m_{I}^{2}),
\end{equation}
where
$v=246 GeV$, is the vacuum expectation value ($vev$) of the Higgs field,  $m_{R}$, $m_{I}$ and $M_{k} (k=1,2,3)$ are the masses of $\sqrt{2} Re[\eta^{0}]$,  $\sqrt{2} Im [\eta^{0}]$ and right-handed neutrino respectively.
\noindent The lepton flavor violating (LFV) processes such as $\mu\to e\gamma$ is induced at one loop level. Thus the branching ratio for $\mu\to e\gamma$ process is given as \cite{Kubo:2006yx,Ma:2001mr}
\begin{equation}
    Br(\mu\to e\gamma)=\frac{3 \alpha_{em}}{64 \pi(G_{F}m_{0}^{2})^{2}}\left|\sum_{k=1}^{3}Y_{\mu k}Y_{e k}^{*}F\left(\frac{M_{k}}{m_{0}}\right)\right|^{2},
\end{equation}
where $\alpha_{em}$ is the fine structure constant for electromagnetic coupling,  $G_{F}$ is  Fermi coupling constant and $F(r)$ is given as
\begin{equation}
F(r)=\frac{1-6r^{2}+3r^{4}+2r^{6}-6r^{4}  \ln r^{2}}{6(1-r^{2})^{4}} , \hspace{0.2cm} r\equiv \frac{M_{k}}{m_{0}}.   
\end{equation}
%In the next section we will discuss co-annihilation of  cold dark matter and relic density of dark matter at freeze-out temperature.

It is interesting to note that this model provides the simultaneous measurement of DM and neutrino mass. Being $Z_2$ odd, the lightest of $N_k$ will be the stable DM candidate. Taking into account the coannihilation effect \cite{Vicente:2014wga}, the predicted cold dark matter abundance as well as the branching ratio of the lepton-flavor-violating (LFV) $\mu\to e \gamma$ process can be consistent with observations, within model at the same time. Here we consider mass of the lightest $N_k$ (DM) is nearly degenerate with the next singlet fermion $N_2$ and the right-handed neutrino mass spectrum considered is  $M_{1}\leq M_{2} < M_{3}$ \cite{Griest}. The product of co-annihilation cross-section and relative velocity of annihilating particle $v_r$ is given by \cite{Suematsu:2009ww}
\begin{equation}
\label{a1}
    \sigma_{ij}|v_{r}|=a_{ij}+b_{ij}v_{r}^{2},
\end{equation}
where

\begin{align}
\left.
 \begin{array}{lll}
     a_{ij}=\frac{1}{8\pi} \frac{M_{1}^2}{(M_{1}^2+m_{0}^{2})^{2}}\sum_{\alpha, \beta}(Y_{\alpha i}Y_{\beta j}-Y_{\alpha j}Y_{\beta i})^{2},\\
    b_{ij}=\frac{m_{0}^{4}-3m_{0}^{2}M_{1}^{2}-M_{1}^{4}}{3(M_{1}^{2}+m_{0}^{2})^{2}} a_{ij}+\frac{1}{12 \pi}\frac{M_{1}^{2}(M_{1}^{4}+m_{0}^{4})}{(M_{1}^{2}+m_{0}^{2})^{4}}\sum_{\alpha,\beta} Y_{\alpha i}Y_{\alpha j}Y_{\beta i}Y_{\beta j},
 \end{array}
 \right\}.
 \end{align}
Here $\sigma_{ij} (i,j=1,2)$ in Eqn. (9) represents  the annihilation cross-section for the process $N_{i} N_{j}\to x x^{'}$, $d M = (M_{2}-M_{1})/M_{1}$ is the mass splitting ratio for the nearly degenerate singlet fermions, $x=M_{1}/T$ i.e. ratio of DM mass to the temperature T.
If $g_{1}$, $g_{2}$ represent the number of degrees of  freedom of singlet fermions $N_{1}$ and $N_{2}$, respectively, the effective cross section is given by
\begin{equation}
\label{a2}
    \sigma_{eff}=\frac{g_{1}^{2}}{g_{eff}^{2}}\sigma_{11}+\frac{2g_{1}g_{2}}{g_{eff}^{2}}\sigma_{12}(1+dM)^{3/2}\exp(-xdM)+\frac{g_{2}^{2}}{g_{eff}^{2}}\sigma_{22}(1+dM )^{3}\exp(-2 xdM),
\end{equation}
\begin{equation}
\label{a3}
    g_{eff}=g_{1}+g_{2}(1+dM)^{3/2} \exp(-xdM),
\end{equation}
with $dM \simeq 0$ ($N_{1}$ is nearly degenerate with $N_{2}$) and using Eqns.(\ref{a1}) and (\ref{a3}) in Eqn.(\ref{a2}) we get
\begin{equation}
\sigma_{eff}|v_{r}|=\left(\frac{\sigma_{11}}{4}+\frac{\sigma_{12}}{2}+\frac{\sigma_{22}}{4}\right)|v_{r}|,
\end{equation}
  \begin{equation*}
      =a_{eff}+b_{eff}v_{r}^{2},  
  \end{equation*}
where 
\begin{align}
    \left.
    \begin{array}{cc}
      a_{eff}=\frac{a_{11}}{4}+\frac{a_{12}}{2}+\frac{a_{22}}{4},\\
          b_{eff}=\frac{b_{11}}{4}+\frac{b_{12}}{2}+\frac{b_{22}}{4},
\end{array}
\right\}.
\end{align}
The thermal average cross section is
\begin{equation}
\label{lebel1}
    <\sigma_{eff}|v_{r}|> = a_{eff}+6 b_{eff}/x,
\end{equation}
 which has  dependence on temperature through the relation $x=M_{1}/T$ and relic density of cold dark matter is given by

\begin{equation}
\label{omega}
    \Omega h^{2} =\frac{1.07\times 10^{9} \text{GeV}^{-1} x_{f}}{(a_{eff}+3 b_{eff}/x_{f}) g_{*}^{1/2} m_{Pl}},
\end{equation}
where $m_{Pl}=1.22\times10^{19}$GeV , $g_{*}=106.75$
and
\begin{equation}
 x_{f}=\ln \frac{0.038 g_{eff} m_{Pl} M_{1}<\sigma_{eff}|v_{r}|> }{g_{*}^{1/2}x_{f}^{1/2} }.   
\end{equation}
Also $x_{f}=M_{1}/T_{f}\approx 25$, $T_{f}$ is the freeze-out temperature\cite{Kolb:1990vq}.

\section{ Dark matter and $0\nu\beta\beta$ decay in the  hybrid textures of neutrino mass matrix}
The general neutrino mass matrix is written as\\
\begin{equation}
M_{\nu}= U diag (m_{1},m_{2}e^{i\alpha_2},m_{3}e^{i\alpha_3}) U^{T}\equiv\begin{pmatrix}
M_{ee} & M_{e\mu} & M_{e\tau}\\
M_{e\mu} & M_{\mu\mu} & M_{\mu\tau} \\
M_{e\tau} & M_{\mu\tau} &  M_{\tau\tau} 
\end{pmatrix},
\end{equation}
 \\
 where $U$ is the Pontecorvo–Maki–Nakagawa–Sakata (PMNS) mixing matrix, ($m_1,m_2,m_3$) are three neutrino mass eigenvalues and ($\alpha_2,\alpha_3$) are Majorana phases. In general, the matrix $U$ is parameterized in term of three mixing angles ($\theta_{12},\theta_{23}, \theta_{13}$) and one Dirac $CP$ violating phase $\delta$. Therefore, the elements of the the neutrino mass matrix are functions of total nine parameters viz.   three mass eigen values ($m_1,m_2,m_3$), three mixing angles ($\theta_{12},\theta_{23}, \theta_{13}$),  two majorana phases ($\alpha_2,\alpha_3$) and one Dirac $CP$ phase $\delta$ as shown below
 \begin{equation}
     M_{\alpha\beta }=f(m_1,m_2,m_3,\theta_{12},\theta_{23},\theta_{13},\alpha_2,\alpha_3,\delta).
 \end{equation}
 Also the $\mu$-$\tau$ sector is expressed as function of $M_{ee}$, $M_{e\mu}$, $M_{e\tau}$ three mixing angles ($\theta_{12},\theta_{23}, \theta_{13}$) and one Dirac $CP$ phase $\delta$ as \cite{Kitabayashi:2015jdj,Kitabayashi:2015tka}
 \begin{equation}
     M_{\mu \tau }=f( M_{ee},M_{e\mu},M_{e\tau},\theta_{12},\theta_{23},\theta_{13},\delta).
 \end{equation}
 For example\\
\begin{align*} 
M_ {\mu\mu} & = (a + 
     b\cos {2\theta_ {23}}) M_ {ee} + ((c + d) s_ {23} + 
     2 s_ {23}^{2} e) M_ {e\mu} + ((c + d) c_ {23} + 
     2 c_ {23}^{2} f) M_ {e\tau}, \\
M_ {\mu\tau} & = (a - 
     b\cos {2\theta_ {23}}) M_ {ee} + ((d - c) s_ {23} - 
     2 s_ {23}^{2} e) M_ {e\mu} + ((d - c) c_ {23} - 
     2 c_ {23}^{2} f) M_ {e\tau}, \\
M_ {\tau\tau} & = -b\sin{2\theta_{23}} M_ {ee} - (s_ {23}\tan {2\theta_ {23}}c + 
     e) M_ {e\mu} - (c_ {23}\tan {2\theta_ {23}}c - f) M_ {e\tau}.
    \end {align*}
where
\begin {align*}
c_ {ij} &\equiv\cos\theta_ {ij}, \\
s_ {ij} &\equiv\sin\theta_ {ij}, \\
a & =  \frac {1} {2} (1 + e^{-2 i\delta}), \\
b & = \frac {1} {2} (1 - e^{-2 i\delta}), \\
c &=-\frac{1}{2}\cos{2\theta_{23}} (e^{-i \delta}\cot{2\theta_{13}}-e^{i\delta}s_{13}c_{13}^{-1}),\\
d &=  e^{-i \delta}\cot{2\theta_{23}}-\frac{1}{2}e^{i\delta}s_{13}c_{13}^{-1},\\
e &=   c_{23}(\cot{2\theta_{12}}\sec{\theta_{13}}-e^{-i\delta}\csc{2\theta_{23}}s_{13}c_{13}^{-1}),\\
f &=s_{23}(\cot{2\theta_{12}}\sec{\theta_{13}}-e^{-i\delta}\csc{2\theta_{23}}s_{13}c_{13}^{-1}),
\end{align*}
($i,j=1,2,3; i<j$) and neutrino mass eigenvalue as
 \begin{equation}
\begin{split}
m_{1,2,3}=f(M_{ee},M_{e\mu},M_{e\tau},\theta_{12},\theta_{23},\theta_{13},\delta,\alpha_{2},\alpha_{3}).
\end{split}
\end{equation}

%In the later study, we find that every element $M_{\alpha\beta}$, $((\alpha,\beta)=e,\mu,\tau)$, of neutrino mass matrix  is proportional to non-zero $|M_{ee}|$.

%After setting out the basic frame work of the scotogenic model and modalities to calculate relic density of DM in earlier sections, here we analyse neutrino mass matrix
\noindent We are interested in lepton number violating ($0\nu\beta\beta$) decay, which also confirms the Majorana nature of neutrinos. Every non-zero element of neutrino mass matrix can be written as proportional to effective Majorana mass $M_{ee}$ as \cite{Kitabayashi:2015jdj,Kitabayashi:2015tka}
\begin{equation}
\label{meq}
M_{\alpha\beta}=f_{\alpha\beta}(\theta_{12},\theta_{23},\theta_{13},\delta) M_{ee},
\end{equation} 
where ($\alpha,\beta=e,\mu,\tau$).

In later study, using Eqn. (22),  we identify nine hybrid textures. From literature we find out of these nine only  six  reproduce low energy phenomenology \cite{Dev:2009he}. Now, we use Eqns. (4) and (24) in Eqn. (22) to incorporate dark matter and $0\nu\beta\beta$ decay in our study. Finally, by using the current experimental bound on relic density of dark matter we find only five hybrid textures which successfully account for the dark matter and $0\nu\beta\beta$ decay \cite{Ankush:2021opd}. These five hybrid textures are as shown below

\noindent 
\begin{center}
\vspace{0.4cm}
$G_{1}:\begin{pmatrix}
M_{e e} & \Delta & \Delta \\
- & 0 & M_{\mu \tau} \\
- & - &  M_{\tau \tau} 
\end{pmatrix}$, 
\hspace{0.4cm}
$G_{2}:\begin{pmatrix}
M_{ee} & 0 & M_{e \tau}\\
- & \Delta & \Delta \\
- & - &  M_{\tau \tau} 
\end{pmatrix}$, \hspace{0.4cm}
$G_{3}:\begin{pmatrix}
M_{e e} & 0 & M_{e \tau}\\
- & M_{\mu \mu} & \Delta \\
- & - & \Delta 
\end{pmatrix}$,\\
\vspace{0.4cm}
$G_{4}:\begin{pmatrix}
M_{e e} & M_{e \mu} & 0 \\
- & \Delta & \Delta \\
- & - &  M_{\tau \tau} 
\end{pmatrix}$, \hspace{0.4cm}
$G_{5}:\begin{pmatrix}
M_{e e} & M_{e\mu} & 0\\
- & M_{\mu \mu} & \Delta \\
- & - &\Delta 
\end{pmatrix}$,
\end{center}

\noindent where $\Delta$ represents the equal elements. In general, the elements of neutrino mass matrix for each texture is written as

\begin{equation}
\label{meq}
M_{\alpha\beta}=f^\mathscr{X}_{\alpha\beta}(\theta_{12},\theta_{23},\theta_{13},\delta) M_{ee},
\end{equation} 
where ($\alpha,\beta=e,\mu,\tau$) and $\mathscr{X}=G_{1....5}$. The constraining equations for each hybrid texture is given in Table \ref{Tab2} and the corresponding expression for the coefficients $f^\mathscr{X}_{\alpha\beta}$ are given in Eqns. (24-28).

% It is interesting to note that the overall scale of neutrino mass is governed by non-zero $M_{ee}$ i.e.

%\noindent Eqn.(\ref{meq}) provide an important link between dark matter and $0\nu\beta\beta$ decay amplitude $\left|M_{ee}\right|$. Using Eqn.(\ref{I}), we calculate $M_{\alpha\beta}$, in Eqn.(\ref{meq}), in terms of loop functions $\Lambda_{k}$ and Yukawa couplings. All the six elements of neutrino mass matrix are calculated and given in Table \ref{Tab2} for each allowed hybrid texture $G_{1....5}$. 

\begin{table}[H]
    \centering
    \begin{tabular}{|c|c|}
    \hline
      Texture   &  Constraining Equations \\
      \hline
        $G_1$      &  $\begin{array} {lcl}Y_{e1}^{2}\Lambda_{1}+Y_{e2}^{2}\Lambda_{2}+Y_{e3}^{2}\Lambda_{3}&=& M_{ee}\\
Y_{e1}Y_{\mu1}\Lambda_{1}+Y_{e2}Y_{\mu2}\Lambda_{2}+Y_{e3}Y_{\mu3}\Lambda_{3}&=& f_{e\mu}^{G_1} M_{ee}\\
Y_{e1}Y_{\tau1}\Lambda_{1}+Y_{e2}Y_{\tau2}\Lambda_{2}+Y_{e3}Y_{\tau3}\Lambda_{3}&=&f_{e\mu}^{G_1} M_{ee}\\
Y_{\mu1}^{2}\Lambda_{1}+Y_{\mu2}^{2}\Lambda_{2}+Y_{\mu3}^{2}\Lambda_{3}&=&0\\
Y_{\mu1}Y_{\tau1}\Lambda_{1}+Y_{\mu2}Y_{\tau2}\Lambda_{2}+Y_{\mu3}Y_{\tau3}\Lambda_{3}&=&f_{\mu\tau}^{G_1} M_{ee}\\
Y_{\tau1}^{2}\Lambda_{1}+Y_{\tau2}^{2}\Lambda_{2}+Y_{\tau3}^{2}\Lambda_{3}&=& f_{\tau\tau}^{G_1} M_{ee} \end{array}$  \\
         \hline

\hline
       $G_2$   &  $\begin{array} {lcl}Y_{e1}^{2}\Lambda_{1}+Y_{e2}^{2}\Lambda_{2}+Y_{e3}^{2}\Lambda_{3}&=& M_{ee}\\
Y_{e1}Y_{\mu1}\Lambda_{1}+Y_{e2}Y_{\mu2}\Lambda_{2}+Y_{e3}Y_{\mu3}\Lambda_{3}&=& 0 \\
Y_{e1}Y_{\tau1}\Lambda_{1}+Y_{e2}Y_{\tau2}\Lambda_{2}+Y_{e3}Y_{\tau3}\Lambda_{3}&=&f_{e\tau}^{G_2} M_{ee}\\
Y_{\mu1}^{2}\Lambda_{1}+Y_{\mu2}^{2}\Lambda_{2}+Y_{\mu3}^{2}\Lambda_{3}&=&f_{\mu\tau}^{G_2} M_{ee}\\
Y_{\mu1}Y_{\tau1}\Lambda_{1}+Y_{\mu2}Y_{\tau2}\Lambda_{2}+Y_{\mu3}Y_{\tau3}\Lambda_{3}&=&f_{\mu\tau}^{G_2} M_{ee}\\
Y_{\tau1}^{2}\Lambda_{1}+Y_{\tau2}^{2}\Lambda_{2}+Y_{\tau3}^{2}\Lambda_{3}&=& f_{\tau\tau}^{G_2} M_{ee}  \end{array}$ \\
      \hline
    $G_3$      & $\begin{array} {lcl}Y_{e1}^{2}\Lambda_{1}+Y_{e2}^{2}\Lambda_{2}+Y_{e3}^{2}\Lambda_{3}&=& M_{ee}\\
Y_{e1}Y_{\mu1}\Lambda_{1}+Y_{e2}Y_{\mu2}\Lambda_{2}+Y_{e3}Y_{\mu3}\Lambda_{3}&=& 0\\
Y_{e1}Y_{\tau1}\Lambda_{1}+Y_{e2}Y_{\tau2}\Lambda_{2}+Y_{e3}Y_{\tau3}\Lambda_{3}&=&f_{e\tau}^{G_3} M_{ee}\\
Y_{\mu1}^{2}\Lambda_{1}+Y_{\mu2}^{2}\Lambda_{2}+Y_{\mu3}^{2}\Lambda_{3}&=&f_{\mu\mu}^{G_3} M_{ee}\\
Y_{\mu1}Y_{\tau1}\Lambda_{1}+Y_{\mu2}Y_{\tau2}\Lambda_{2}+Y_{\mu3}Y_{\tau3}\Lambda_{3}&=&f_{\mu\tau}^{G_3} M_{ee}\\
Y_{\tau1}^{2}\Lambda_{1}+Y_{\tau2}^{2}\Lambda_{2}+Y_{\tau3}^{2}\Lambda_{3}&=& f_{\mu\tau}^{G_3} M_{ee} \end{array} $  \\
         \hline
              
    $G_4$     &   $\begin{array} {lcl}Y_{e1}^{2}\Lambda_{1}+Y_{e2}^{2}\Lambda_{2}+Y_{e3}^{2}\Lambda_{3}&=& M_{ee}\\
 Y_{e1}Y_{\mu1}\Lambda_{1}+Y_{e2}Y_{\mu2}\Lambda_{2}+Y_{e3}Y_{\mu3}\Lambda_{3}&=& f_{e\mu}^{G_4} M_{ee}\\
Y_{e1}Y_{\tau1}\Lambda_{1}+Y_{e2}Y_{\tau2}\Lambda_{2}+Y_{e3}Y_{\tau3}\Lambda_{3}&=&0\\
Y_{\mu1}^{2}\Lambda_{1}+Y_{\mu2}^{2}\Lambda_{2}+Y_{\mu3}^{2}\Lambda_{3}&=&f_{\mu\mu}^{G_4} M_{ee}\\
Y_{\mu1}Y_{\tau1}\Lambda_{1}+Y_{\mu2}Y_{\tau2}\Lambda_{2}+Y_{\mu3}Y_{\tau3}\Lambda_{3}&=&f_{\mu\mu}^{G_4} M_{ee}\\
Y_{\tau1}^{2}\Lambda_{1}+Y_{\tau2}^{2}\Lambda_{2}+Y_{\tau3}^{2}\Lambda_{3}&=& f_{\tau\tau}^{G_4} M_{ee}  \end{array}$ \\
         \hline
               
   $G_5$      &  $\begin{array} {lcl}Y_{e1}^{2}\Lambda_{1}+Y_{e2}^{2}\Lambda_{2}+Y_{e3}^{2}\Lambda_{3}&=& M_{ee}\\
Y_{e1}Y_{\mu1}\Lambda_{1}+Y_{e2}Y_{\mu2}\Lambda_{2}+Y_{e3}Y_{\mu3}\Lambda_{3}&=& f_{e\mu}^{G_5} M_{ee}\\
Y_{e1}Y_{\tau1}\Lambda_{1}+Y_{e2}Y_{\tau2}\Lambda_{2}+Y_{e3}Y_{\tau3}\Lambda_{3}&=&0\\
Y_{\mu1}^{2}\Lambda_{1}+Y_{\mu2}^{2}\Lambda_{2}+Y_{\mu3}^{2}\Lambda_{3}&=&f_{\mu\mu}^{G_5} M_{ee}\\
Y_{\mu1}Y_{\tau1}\Lambda_{1}+Y_{\mu2}Y_{\tau2}\Lambda_{2}+Y_{\mu3}Y_{\tau3}\Lambda_{3}&=&f_{\mu\tau}^{G_5} M_{ee}\\
Y_{\tau1}^{2}\Lambda_{1}+Y_{\tau2}^{2}\Lambda_{2}+Y_{\tau3}^{2}\Lambda_{3}&=&f_{\mu\tau}^{G_{3}} M_{ee} \end{array} $  \\
         \hline
  
    \end{tabular}
\caption{Constraining equations relating loop factors and Yukawa couplings to the effective Majorana neutrino mass $\left|M_{ee}\right|$ for all hybrid textures $G_{1....5}$.}
\label{Tab2}
\end{table}
%\noindent The corresponding expressions of $f^X_{\alpha\beta}$ can be read from Eqns.(26-30). Also $f_{ee}$ coefficients for all textures are unity.  
%\noindent The coefficients, $f_{\alpha \beta}^{X}$ for all textures are as follows\\

\begin{align}
\left.
 \begin{array}{lll}
f_{e\mu}^{G_{1}}&=&\frac{\sin{2\theta_{12}} \sin{2\theta_{13}}(e^{2i\delta}c_{23}^{2}+s^{2}_{23})}{-4e^{2i\delta}L_{5}(c_{23}-s_{23}) c^{2}_{23} K_{3}+\frac{e^{i\delta}}{\csc{2\theta_{12}}} (2 e^{2i\delta}c_{23}^{2} K_{3}+s_{23}(K_{1}-\frac{2 M_{3}}{\sin{2\theta_{12}}}))},\\
f_{\mu\tau}^{G_{1}}&=&\frac{\sqrt{2}\cos(\frac{\pi}{4}+\theta_{23})(\frac{-4e^{i\delta} L_{5}}{\sec{2\theta_{23}}}+\frac{2\sin{2\theta_{12}}}{\csc{2\theta_{23}}}( s_{13}^{2} +e^{2i\delta}(\frac{\cos^{2}{\theta_{13}}}{\sin{2\theta_{23}}}+1)))}{-4e^{3i\delta}L_{5}(c_{23}-s_{23}) c^{2}_{23} K_{3}+\frac{e^{i\delta}}{\csc{2\theta_{12}}}  (2 e^{2i\delta}c_{23}^{2} K_{3}+s_{23}(K_{1}-\frac{2 M_{3}}{\sin{2\theta_{12}}}))},\\
f_{\tau\tau}^{G_{1}}&=&\frac{\sin{2\theta_{12}}(  \frac{2(c_{23}-s_{23})}{\csc^{2}{\theta_{13}}\csc^{2}{\theta_{23}}}+\frac{e^{2i\delta}}{\sec{\theta_{23}}}(\frac{M_{3}}{\sin{2\theta_{12}}}+\sin{2\theta_{23}})-e^{i\delta}L_{5}(L_{1}-K_{5}))}{-4e^{3i\delta}L_{5}(c_{23}-s_{23}) c^{2}_{23} K_{3}+\frac{e^{i\delta}}{\csc{2\theta_{12}}} (2 e^{2i\delta}c_{23}^{2} K_{3}+s_{23}(K_{1}-\frac{2 M_{3}}{\sin{2\theta_{12}}}))}, 
\end{array}
\right\}
\end{align}
%and for texture  T$6$, we obtain the following expressions

%\hspace{0.35cm} $ f_{\tau \tau}^{T6}=0$ and $ f_{e \mu}^{T6}=f_{e \tau}^{T6}$

\begin{align}
\left.
 \begin{array}{lll}
f_{e\tau}^{G_{2}}&=&\frac{\sin{2\theta_{12}} \sin{2\theta_{13}} (K_{1}+e^{2i\delta}K_{2})}{2e^{i\delta} \sin{2\theta_{12}} (2e^{2i\delta} c_{23}^2 K_{3}+s_{23}(K_{4}-\sin{2\theta_{23}}))+2e^{2i\delta}L_{5}(K_{5}+L_{1})},\\
f_{\mu\tau}^{G_{2}}&=&\frac{4s_{23} (e^{i\delta} c_{23} s_{12}+c_{12} s_{13} s_{23})(e^{i\delta}c_{12} c_{23}-s_{12} s_{13} s_{23})}{2e^{i\delta} \sin{2\theta_{12}} (2e^{2i\delta} c_{23}^2 K_{3}+s_{23}(K_{4}-\sin{2\theta_{23}}))+e^{3i\delta}L_{5}(K_{5}+L_{1})},\\
f_{\tau \tau}^{G_{2}}&=&\frac{ \frac{(N_{1}-2K_{5})}{\csc{2\theta_{12}} \csc_{13}^{2}}+ \frac{e^{2i\delta}}{\csc{2\theta_{12}}}  (2( \cos{3\theta_{23}} +\frac{K_{3} \cos{2\theta_{13}}}{s_{13}^{2}}+e^{i\delta}L_{5}(K_{5}+2N_{1}))-N_{1})}{2e^{i\delta} \sin{2\theta_{12}} (2e^{2i\delta} c_{23}^2 K_{3}+s_{23}(K_{4}-\sin{2\theta_{23}}))+2e^{3i\delta}L_{5}(K_{5}+L_{1})},
\end{array}
 \right\}
 \end{align}

%   for texture  T$2$, we obtain the following expressions
 
%\hspace{0.35cm} $f_{e\mu}^{T2}=0$ and $f_{\mu \tau}^{T2}=f_{\tau \tau}^{T2}$

\begin{align}
\left.
 \begin{array}{lll}
f_{e\tau}^{G_{3}}&=&\frac{\sin{2\theta_{12}} s_{13} ((K_{1}+2 \cos{2\theta_{23}})+e^{2i\delta}(K_{2}-2 \cos{2\theta_{23}}))}{8e^{2i\delta}L_{5}s^{2}_{23}K_{3}-e^{i\delta} \sin{2\theta_{12}}(L_{3}-2 K_{4} s_{23})+e^{3i\delta}\sin{2\theta_{12}} s_{13}^{2}(K_{5}+N_{1})},\\
f_{\mu \tau}^{G_{3}}&=&\frac{2s_{23}(e^{2i\delta} K_{4} \sin{2\theta_{12}}+ 2 \cos^{2}{\theta_{23}} L_{4}+2e^{i\delta}\sin{2\theta_{23}} L_{5})}{8e^{3i\delta}L_{5}s^{2}_{23}K_{3}-e^{2i\delta} \sin{2\theta_{12}}(L_{3}-2 K_{4} s_{23})+e^{4i\delta}\sin{2\theta_{12}} s_{13}^{2}(K_{5}+N_{1})},\\
f_{\mu \mu }^{G_{3}}&=&\frac{\frac{2}{\csc{2\theta_{12}}}(\frac{2(2 c_{23}-s_{23})}{\csc{\theta_{13}}^{2}}-\frac{e^{2i\delta}}{\sec{\theta{23}}} (K_{4}-K_{1}-\frac{2}{\csc{2\theta_{23}}})+e^{i\delta}\frac{ s_{13}}{\tan{2\theta_{12}}}(K_{5}-2N_{1}))}{8e^{3i\delta}L_{5}s^{2}_{23}K_{3}-e^{2i\delta} \sin{2\theta_{12}}(L_{3}-2 K_{4} s_{23})+e^{4i\delta}\sin{2\theta_{12}} s_{13}^{2}(K_{5}+N_{1})},
\end{array}
\right\}
\end{align}

%for texture  T$3$, we obtain the following expressions

%\hspace{0.35cm}$ f_{e\tau}^{T3}=0$ and $ f_{\mu \mu}^{T3}=f_{\tau \tau}^{T3}$
\begin{align}
\left.
 \begin{array}{lll}
f_{e\mu}^{G_{4}}&=&\frac{\sin{2\theta_{12}} \sin{2\theta_{13}}(K_{1}+e^{2i\delta}K_{2})}{\frac{-8e^{2i\delta} c_{13}^2}{s_{13^2}} K_{3} L_{5}+\frac{2e^{i\delta}}{\csc{2\theta_{12}}}( M_{1}\cos{2\theta_{13}} +M_{2} c_{23} )+e^{3i\delta} L_{4} (K_{5}+L_{1})},\\
f_{\mu\tau}^{G_{4}}&=&\frac{2 c_{23}(e^{2i \delta} M_{3} +2 M_{4} - 2 e^{i \delta} L_{5} \sin{2\theta_{23}})}{\frac{-8e^{3i\delta} c_{13}^2}{s_{13^2}} K_{3} L_{5}+\frac{2e^{2i\delta}}{\csc{2\theta_{12}}}( M_{1}\cos{2\theta_{13}} +M_{2} c_{23} )+e^{4i\delta} L_{4} (K_{5}+L_{1})},\\
f_{\tau\tau}^{G_{4}}&=&\frac{2(\sin{2\theta_{12}}(-2c_{23}^{2} s_{13}^{2} M_{1}+e^{2i\delta}N_{2}-e^{id} L_{5}(2L_{2}+L_{1})))}{\frac{-8e^{3i\delta} c_{13}^2}{s_{13^2}} K_{3} L_{5}+\frac{2e^{2i\delta}}{\csc{2\theta_{12}}}( M_{1}\cos{2\theta_{13}} +M_{2} c_{23} )+e^{4i\delta} L_{4} (K_{5}+L_{1})},  
\end{array}
\right\}
\end{align}

%for texture  T$4$, we obtain the following expressions
 
%\hspace{0.35cm}$ f_{e\tau}^{T4}=0$ and $ f_{\mu \tau}^{T4}=f_{\tau \tau}^{T4}$
\begin{align}
\left.
 \begin{array}{lll}
f_{e\mu}^{G_{5}}&=&\frac{\sin{2\theta_{12}} s_{13}((K_{1}+2 \cos{2\theta_{23}})+e^{2i\delta}(K_{1}+2 \sin{2\theta_{23}}))}{\frac{4e^{3i\delta} M_{4} K_{3}}{s_{13}^{2}}-2e^{i\delta}c_{23} \sin{2\theta_{12}}(K_{2}-c_{13}^{2})-2e^{2i\delta}L_{5}(K_{5}+L_{1})},\\
f_{\mu\mu}^{G_{5}}&=&\frac{16e^{i\delta}L_{5}(L_{2}- \frac{c_{23}}{\csc{2\theta_{23}}})-\frac{4e^{2i\delta}}{\csc{2\theta_{12}}}(\frac{e^{-2i\delta}N_{3}}{ \csc{\theta_{13}}^{2}}-(L_{2}-\frac{2 \cos{2\theta_{13}}K_{3}}{s_{13}^{2}} +\frac{2 }{\csc{3\theta_{23}}}))}{\frac{16 e^{4i\delta} M_{4} K_{3}}{s_{13}^{2}}+8e^{2i\delta}c_{23} \sin{2\theta_{12}}(K_{2}-c_{13}^{2})-8e^{3i\delta}L_{5}(K_{5}+L_{1})},\\
 f_{\mu\tau}^{G_{5}}&=&\frac{4 c_{23}(c_{23} s_{12} s_{13}+e^{i\delta}\frac{\sin{2\theta_{12}}}{2})(-c_{12} c_{23} s_{13}+e^{i\delta}s_{12} s_{23})}{\frac{2 e^{4i\delta}M_{4} K_{3}}{s_{13}^{2}}-e^{2i\delta}c_{23} \sin{2\theta_{12}}(K_{2}-c_{13}^{2})-3e^{3i\delta}L_{5}(K_{5}+L_{1})},
\end{array}
\right\}
\end{align}

where 
\begin{align}
\left.
 \begin{array}{lllllllllllllllllllll}
K_{1}=1-\cos{2 \theta_{23}}-\sin{2 \theta_{23}},\\
K_{2}=1+\cos{2 \theta_{23}}+\sin{2 \theta_{23}},\\
K_{3}=s_{13}^{2}(c_{23}+s_{23}),\\
K_{4}=\cos{2 \theta_{13}}+\cos{2 \theta_{23}},\\
K_{5}=c_{23}-\cos{3 \theta_{23}},\\
L_{1}=s_{23}+\sin{3 \theta_{23}},\\
L_{2}=c_{23}+\cos{3 \theta_{23}},\\
L_{3}=(-1+4 \cos{2 \theta_{13}})\cos{2 \theta_{23}}+\cos{3 \theta_{23}},\\
L_{4}=\sin{2 \theta_{12}} s_{13}^{2},\\
L_{5}=\cos{2 \theta_{12}} s_{13}, \\
M_{1}=c_{23}-2 s_{23},\\
M_{2}=-\cos{2 \theta_{23}}+\sin{2 \theta_{23}},\\
M_{3}=\sin{2 \theta_{12}}(\cos{2 \theta_{13}}-\cos{2 \theta_{23}}),\\
M_{4}=\sin{2 \theta_{12}} s_{13}^{2} s^{2}_{23},\\
%C_{5}=-1+\cos4{ \theta_{23}}+\sin4{ \theta_{23}}
%-2\cos{2 \theta_{13}}(-2+2\cos{2 \theta_{23}}+\sin{2 \theta_{23}}),\\
N_{1}=-s_{23}+\sin{3 \theta_{23}},\\
N_{2}=s_{23}(2(\cos{2 \theta_{23}}+s_{13}^{2})+\sin{2 \theta_{23}}),\\
N_{3}=\cos{3 \theta_{23}}+c_{23}(-1+4\sin{2 \theta_{23}}),\\
\end{array}
\right\}
\end{align}
%\begin{align}
%\left.
% \begin{array}{lll}
%\sin^{2}{\theta_{12}} &= 0.310,\\
%     \sin^{2}{\theta_{13}}&=0.0224 ,\\
%      \sin^{2}{\theta_{23}}&=0.441,
%\end{array}
%\right\}
%\end{align}
  \begin{table}[H]
      \centering
\begin{tabular}{ |c|c|c|c| } 
\hline
Mixing angles & bfp $\pm$ 1$\sigma$ & 3$\sigma$ range \\
\hline

$\theta_{12}/^{o}$ & $33.82^{+0.78}_{-0.76}$ & $31.61\rightarrow36.27$ \\ 
\hline
$\theta_{23}/^{o}$ & $49.6^{+1.0}_{1.2}$ & $40.3\rightarrow52.4$ \\ 
\hline

$\theta_{13}/^{o}$ & $8.61^{+0.13}_{-0.13}$ & $8.22\rightarrow8.99$ \\ 
\hline

$\delta/^{o}$ & $215^{+40}_{-29}$ & $125\rightarrow392$ \\ 
\hline
\end{tabular}
\caption{Global fit data of neutrino mixing angles and CP phase $\delta$ \cite{Esteban:2018azc}.}
\label{tab33}
 \end{table}
\noindent  Clearly, $f_{ee}$ coefficients for all these hybrid textures are unity. All these coefficients $f_{\alpha \beta}^{\mathscr{X}}$  are calculated by randomly generating the mixing angles and CP phase within their allowed range using the data given in Table \ref{tab33}\cite{Esteban:2018azc}.

\begin{section}{Lepton Asymmetry}
Leptogenesis offers a minimal framework to comprehend the dynamical origin of baryon asymmetry of universe, even though the origins of this baryon asymmetry may be completely distinct. In our universe, which most likely started off as baryon symmetric, there are three fundamental conditions called Sakharov's conditions, that must be met for the dynamical generation of this asymmetry. These three conditions are (i) baryon number violation, (ii)  $C$ and $CP$, violation and (iii) departure from thermal equilibrium.\\
In our work, we take into account the out-of-equilibrium decay of the heavy right-handed neutrinos, the eminent method to find the lepton asymmetry \cite{Gautam:2022jrg,Weinberg:1979bt,Kolb:1979qa,Hugle:2018qbw,Borah:2018rca,Racker:2013lua}. The lepton asymmetry due to the decay of the right-handed neutrinos into Higgs and leptons is given by 

\begin{equation}
    \epsilon_{N k}=\sum_i \frac{\Gamma(N_k\rightarrow{L_i+H^\dagger})-\Gamma(N_k\rightarrow{\bar{L_i}+H})}{\Gamma(N_k\rightarrow{L_i+H^\dagger})+\Gamma(N_k\rightarrow{\bar{L_i}+H})}
\end{equation}
At very high temperatures $(T\geq10^{12} GeV)$, all charged flavours are out of equilibrium and behave similarly, resulting in the unflavored regime. However, at temperatures  $T< 10^{12} GeV$ ($T< 10^9 GeV$), tau (muon) Yukawa couplings enter equilibrium, and flavour effects become important in the calculation of lepton asymmetry \cite{Barbieri:1999ma,Abada:2006fw,Abada:2006ea,BhupalDev:2014pfm}. The temperature ranges  $10^9 GeV < T < 10^{12} GeV$  and $T< 10^{9} GeV$  correspond to two and three flavour leptogenesis regimes, respectively.
We assume a hierarchical spectrum for the heavy right handed neutrino, $M_1\leq M_2<M_3$. As a result, the source of lepton asymmetry is the decay of the lightest right handed neutrino $N_1$. The asymmetry caused by the decay of $N_1$ is calculated as \cite{Joshipura:2001ya}

\begin{equation}
   \begin{aligned}
        \epsilon_1^\alpha={} &\frac{1}{8 \pi }\frac{1}{[Y^\dagger Y]_{11}}\sum_{j=2,3} Im[Y^\dagger_{\alpha 1}(Y^\dagger Y)^2_{1j}Y^\dagger_{\alpha j}]g(x_j)\\
    & +\frac{1}{8\pi } \frac{1}{[Y^\dagger Y]_{11}}\sum_{j=2,3} Im[Y^\dagger_{\alpha 1}(Y^\dagger Y)^2_{j1}Y^\dagger_{\alpha j}]\frac{1}{1-x_j}
   \end{aligned}
\end{equation}
where\\
\begin{equation*}
    g(x)=\sqrt{x}\left(1+\frac{1}{1-x}-(1+x)ln\frac{1+x}{x}\right), \text{and}\hspace{0.4cm} x_j=M_j^2/M_1^2.
\end{equation*}
When summed over all flavors ($\alpha=e,\mu,\tau$), the second term in $\epsilon_1^\alpha$ vanishes and sum over all flavors is given by
\begin{equation}
    \epsilon_1=\frac{1}{8 \pi }\frac{1}{[Y^\dagger Y]_{11}}\sum_{j=2,3} Im[(Y^\dagger Y)^2_{1j}]g(x_j)
\end{equation}
The lepton asymmetry through electroweak sphaleron process gives rise to baryon asymmetry is given by \cite{Kuzmin.1} 
\begin{equation}
    Y_B=ck\frac{\epsilon_1}{g_*}
\end{equation}
 where $c$ is the fraction of lepton asymmetry converted to baryon asymmetry, which is approximately equal to -$0.55$, and $k$ is the dilution factor due to washout processes that erase the produced asymmetry, which can be parameterized as \cite{EWKOLB1,Pilaftsis:1998pd,Flanz:1998kr}
  \begin{equation}
    \begin{aligned}
        - k\approx{}& \sqrt{0.1 K} \exp{[-4/(3(0.1 K)^{0.25})]}, \hspace{0.3cm} \text{for} \hspace{0.2cm} K \ge 10^6\\
        &\approx \frac{0.3}{K(ln K)^{0.6}}, \hspace{0.3cm} \text{for} \hspace{0.2cm} 10\leq K \leq 10^6\\
          &\approx \frac{1}{2\sqrt{K^2+9}}, \hspace{0.3cm} \text{for} \hspace{0.2cm}0\leq K \leq 10,
    \end{aligned}
  \end{equation} 
  where K is given by
  \begin{equation}
      K=\frac{\Gamma_1}{H(T=M_1)}=\frac{(Y^{\dagger}Y)_{11}M_1}{8\pi v^2}\frac{M_{pl}}{1.66\sqrt{g_*}M_1^2}.
  \end{equation}
  Here $\Gamma_1$ is the decay width of $N_1$ and $H(T= M1)$
is the
Hubble constant at temperature $(T= M1)$. The factor $g_*$ is
the effective number of relativistic degrees of freedom at
$(T= M1)$ and is approximately equal to 110.\\
 The two flavor and three flavor leptogenesis is given by 
{\begin{equation}
    Y_B^{2 flavor}=\frac{-12}{37 g^*}\left[\epsilon_2 \eta\left(\frac{417}{589}\Tilde{m_2}\right)+\epsilon_1^\tau \eta\left(\frac{390}{589}\Tilde{m_\tau}\right)\right]
\end{equation}
\begin{equation}
    Y_B^{3 flavor}=\frac{-12}{37 g^*}\left[\epsilon_1^e \eta\left(\frac{151}{179}\Tilde{m_e}\right)+\epsilon_1^\mu \eta\left(\frac{344}{537}\Tilde{m_\mu}\right)+\epsilon_1^\tau \eta\left(\frac{344}{537}\Tilde{m_\tau}\right)\right]
\end{equation}}
  Where
\begin{align}
\left.
 \begin{array}{lll}
\epsilon_2&=&\epsilon_1^e+\epsilon_1^\mu,\\ 
\Tilde{m_2}&=&\Tilde{m_e}+\Tilde{m_\mu},\\
\Tilde{m_\alpha}&=&\frac{v^2(Y^\dagger Y)_{\alpha1}}{M_1},
\end{array}
\right\}
\end{align}
   The function $\eta$ is given by
    \begin{equation}
        \eta(\Tilde{m_\alpha})=\left[\left(\frac{\Tilde{m_\alpha}}{8.25\times 10^{-3} eV}\right)^{-1} +\left(\frac{0.2\times 10^-3 eV}{\Tilde{m_\alpha}}\right)^{-1.16}\right]^{-1}
    \end{equation}  

\end{section}
\section{Numerical analysis and discussion}
In the preceding sections, we first set the framework  to find the branching ratio for the flavor violating process $Br(\mu\to e\gamma)$, and relic density of dark matter ($\Omega h^{2}$). We also discuss the expressions which relates dark matter and neutrinoless double beta decay. Then, finally in the last section we discuss the framework to calculate the lepton asymmetry and baryon asymmetry of universe.

\noindent We calculate the loop functions $\Lambda_k$ for each texture using Eqn.(5) by randomly generating quartic coupling $\lambda$, lightest right-handed neutrino mass $M_1$ with in their specified ranges given in Table \ref{tab3}. As stated before, we assume mass hierarchy $M_1\leq M_2 <M_3$  and  $m_0\gtrsim M_1$ in the calculation of $\lambda_k$. Substituting  $\Lambda_k$'s and randomly varying the diagonal Yukawa couplings  ($Y_{e1},Y_{\mu 2},Y_{\tau 3}$) on the left-hand side and $\left|M_{ee}\right|$  in the range ($0-0.25$) eV on the right-hand side of constraining equations given in Table \ref{Tab2}, we calculate the off-diagonal Yukawa couplings and  constraining them in the range $0$ - $1.5$. Furthermore, the numerical analysis imposes bounds on the LFV  process $\mu\rightarrow e\gamma$ i.e., $Br(\mu\rightarrow e\gamma) \leq 4.2\times 10^{-13}$ and on  dark matter mass $M_1$ given in Table \ref{tab4} \cite{MEG:2016leq}.\\
The data obtained in this manner met all of the requirements for the simultaneous validation of dark matter and neutrinoless double beta decay. Now, we use  these masses of the right handed neutrinos and Yukawa couplings in Eqns. (32) and (34) to calculate the lepton asymmetry $\epsilon_1$ and diluation factor $k$ which then put in Eqn. (33) to get the final baryon asymmetry.
 \begin{table}[H]
      \centering
      \begin{tabular}{|c|c|}
      \hline
       Parameter    & Range \\ \hline
        $\lambda$   & ($3-4$)$\times 10^{-9}$\\
        \hline
        $Y_{e1},Y_{\mu2},Y_{\tau3}$ &  $0-1.5$
        \\
        \hline
        $M_1$ & $\mathcal{O}(TeV)$\\
        \hline
      \end{tabular}
      \caption{Ranges of parameters used in the numerical analysis.}
      \label{tab3}
  \end{table}
  \begin{figure}
  \centering
   \includegraphics[scale=0.75]{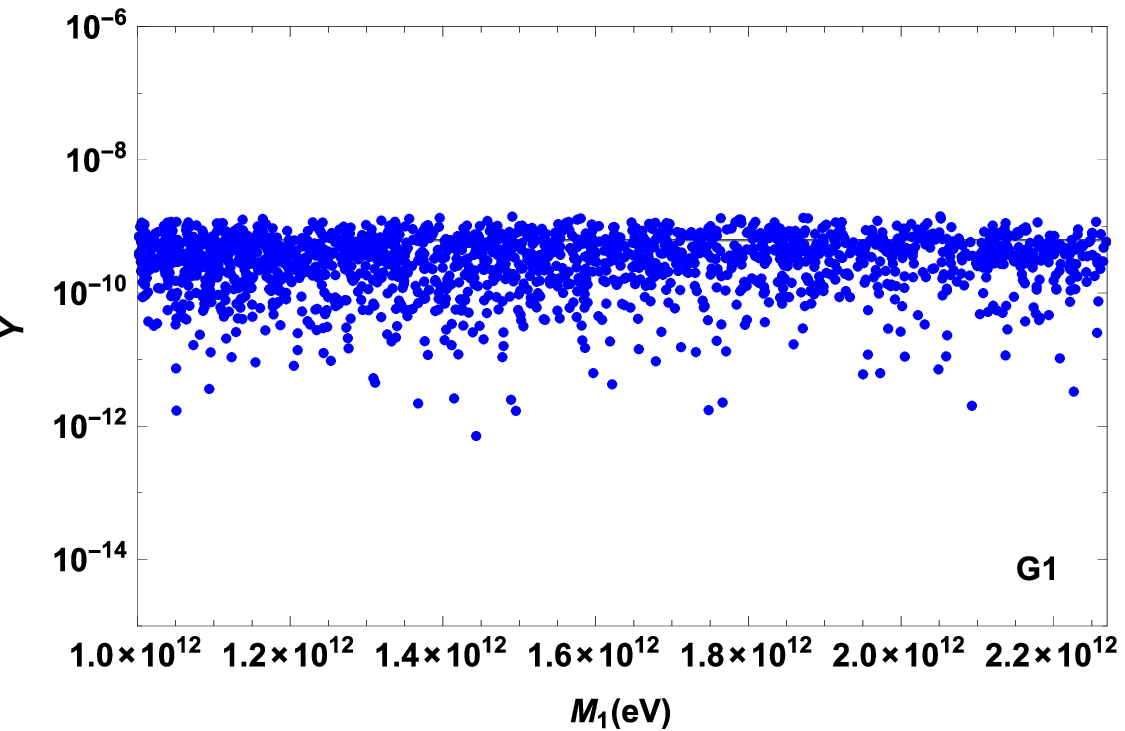}
\caption{Correlation between baryon asymmetry of universe $Y$ and DM mass $M_1$ for texture $G_1$. The horizontal line is the observed value of baryon asymmetry of universe $Y=(6.04\pm 0.08) \times 10^{-10}$ \cite{Planck:2018vyg} .}
\label{Fig1}
\end{figure}

\begin{figure}
\centering
   \includegraphics[scale=0.75]{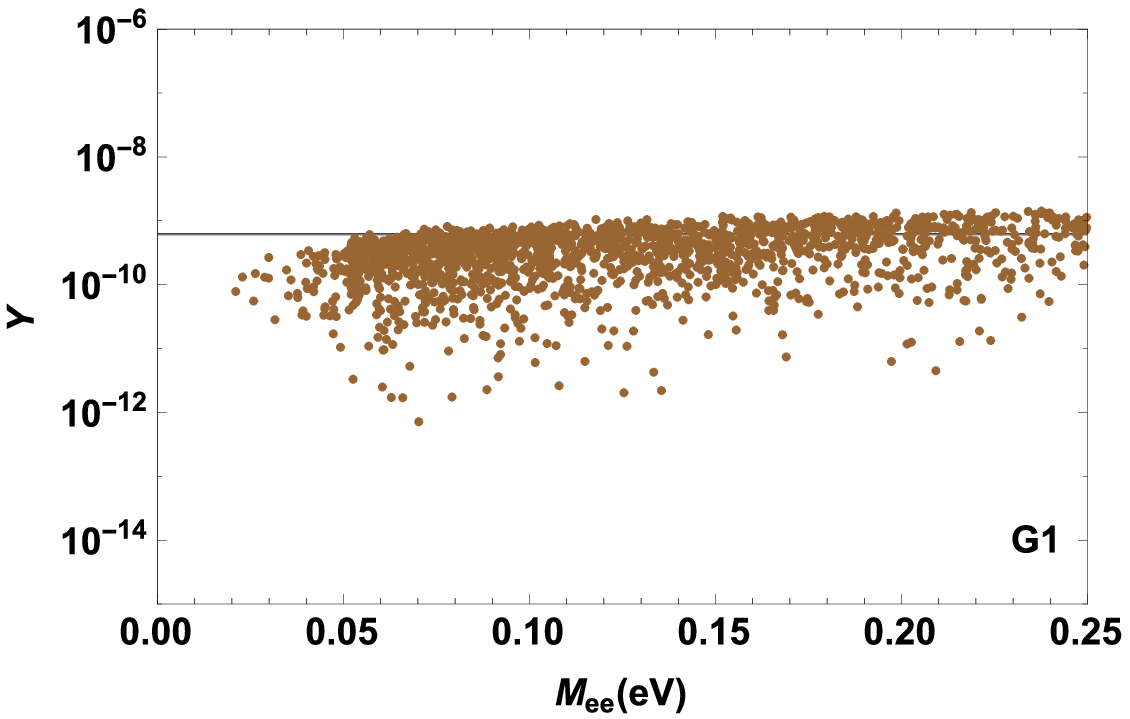}
\caption{Correlation between baryon asymmetry of universe $Y$ with effective Majorana mass $\left|M_{ee}\right|$ for texture $G_1$. The horizontal line is the observed value of baryon asymmetry of universe $Y=(6.04\pm 0.08) \times 10^{-10}$ \cite{Planck:2018vyg}.}
\label{Fig2}
\end{figure}
\begin{figure}
\centering
   \includegraphics[scale=0.75]{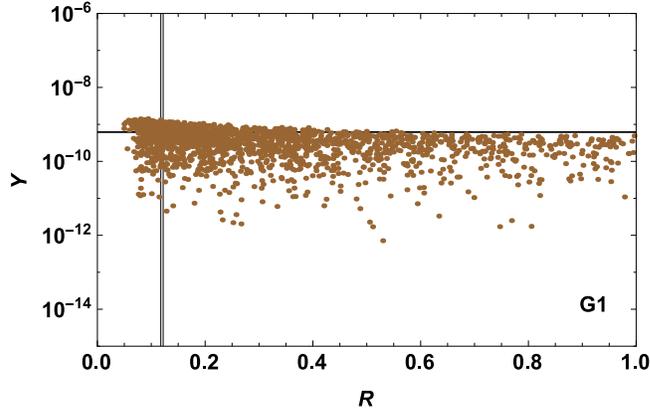}
\caption{Correlation between baryon asymmetry of universe $Y$ and Relic density of DM($\Omega h^2$) for the texture $G_1$. The vertical line is the observed value of relic density of DM $\Omega h^2=0.120\pm 0.001$ and the horizontal line is the observed value of baryon asymmetry of universe $Y=(6.04\pm 0.08) \times 10^{-10}$ \cite{Planck:2018vyg}.}
\label{Fig3}
\end{figure}

\begin{figure}
\centering
   \includegraphics[scale=0.45]{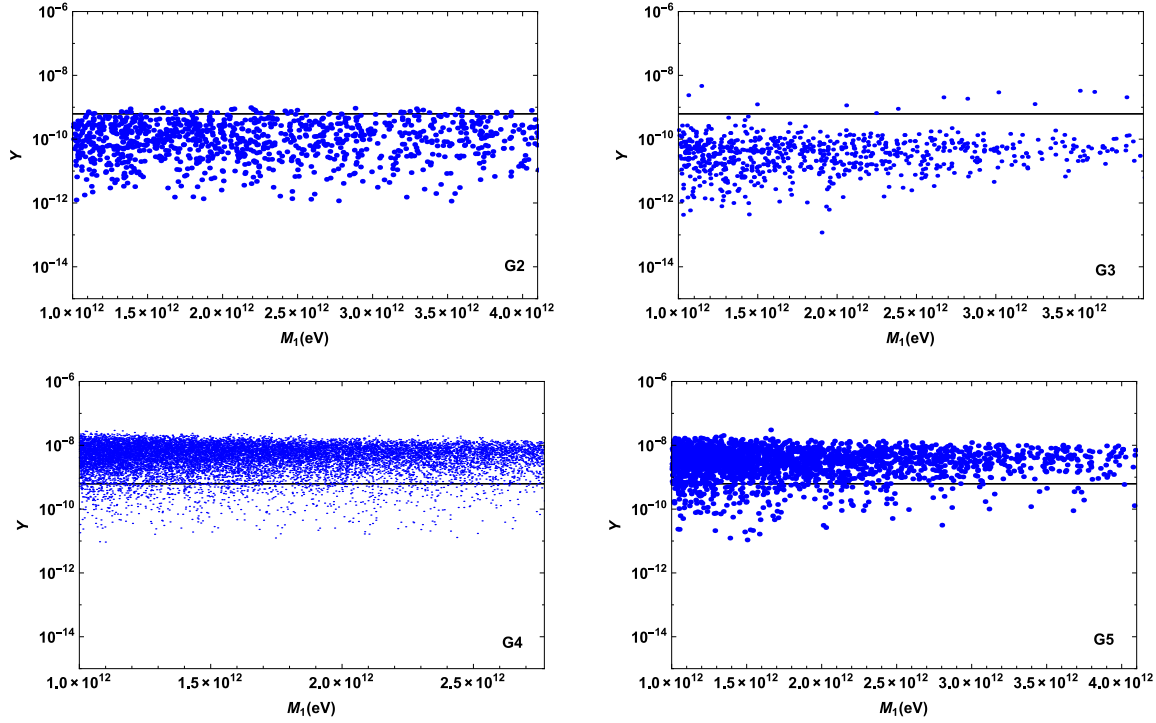}
\caption{Correlation between baryon asymmetry of universe $Y$ and DM mass $M_1$ for textures $G_{2,..,5}$. The horizontal line is the observed value of baryon asymmetry of universe $Y=(6.04\pm 0.08) \times 10^{-10}$ \cite{Planck:2018vyg}.}
\label{Fig4}
\end{figure}

\begin{figure}
\centering
   \includegraphics[scale=0.45]{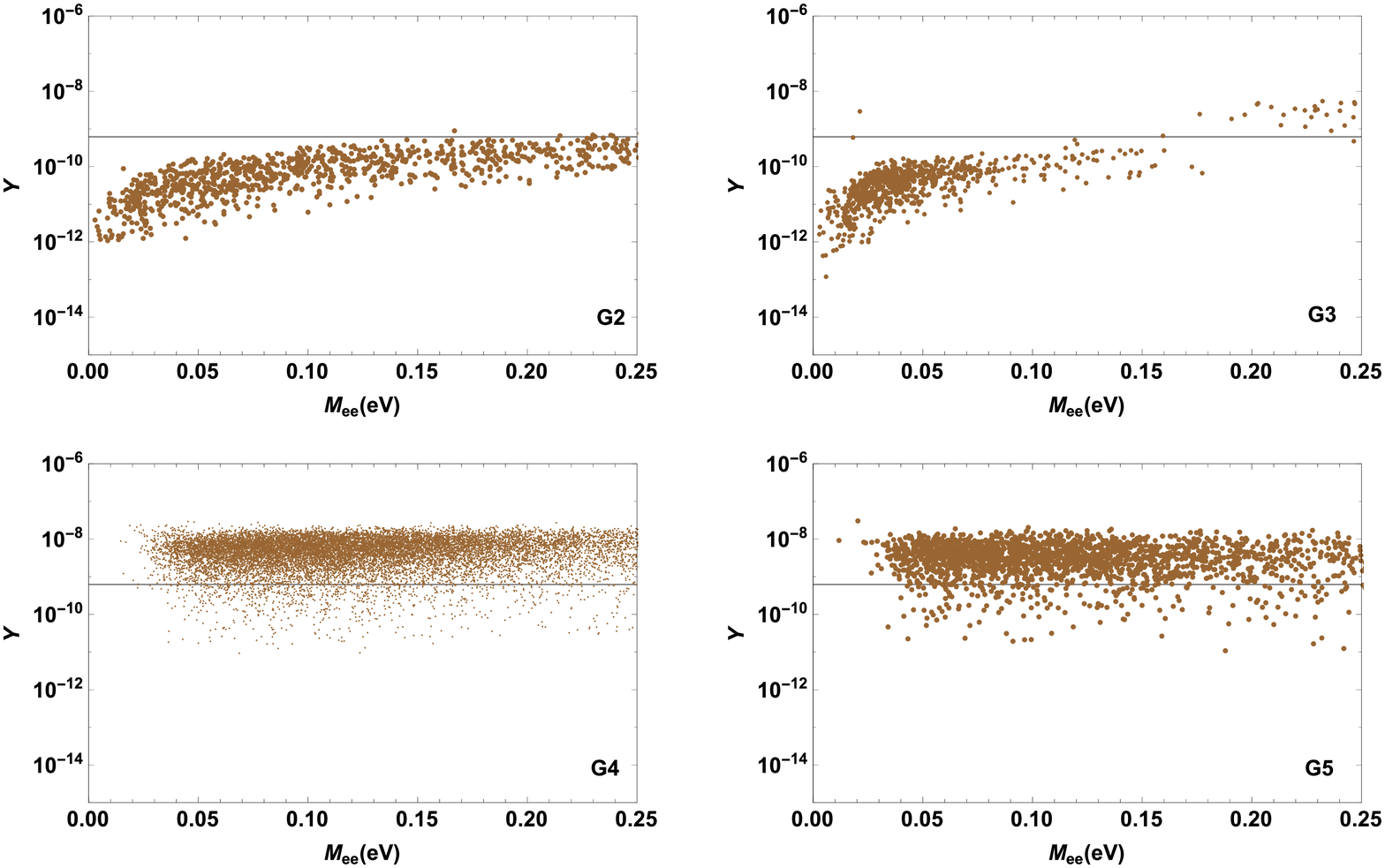}
\caption{Correlation between baryon asymmetry of universe $Y$ with effective Majorana mass $\left|M_{ee}\right|$ for textures $G_{2,..,5}$. The horizontal line is the observed value of baryon asymmetry of universe $Y=(6.04\pm 0.08) \times 10^{-10}$ \cite{Planck:2018vyg}.}
\label{Fig5}
\end{figure}

\begin{figure}
\centering
   \includegraphics[scale=0.45]{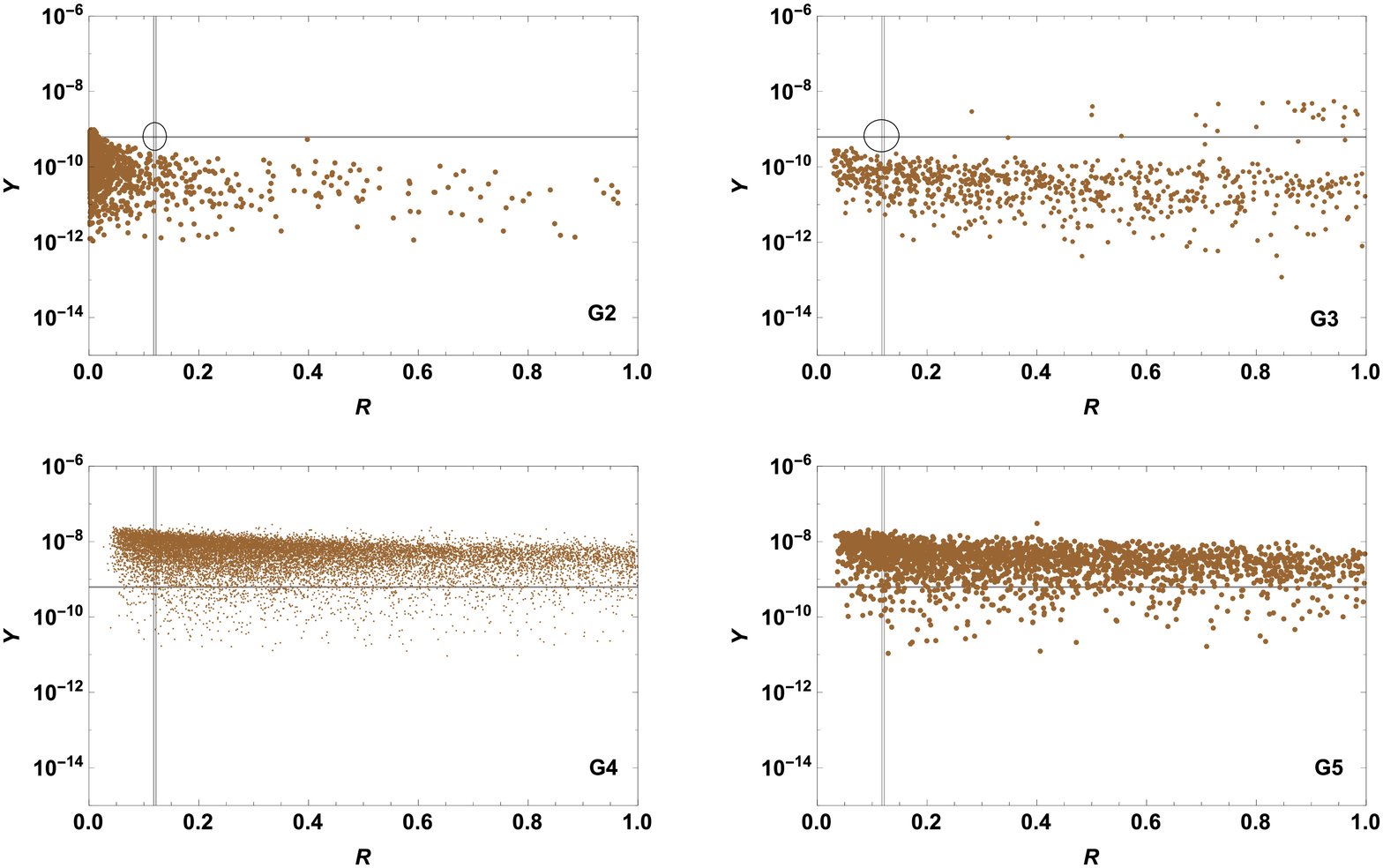}
\caption{Correlation between baryon asymmetry of universe $Y$ and Relic density of DM($\Omega h^2$). The vertical line is the observed value of relic density of DM $\Omega h^2=0.120\pm 0.001$ and the horizontal line is the observed value of baryon asymmetry of universe $Y=(6.04\pm 0.08) \times 10^{-10}$ \cite{Planck:2018vyg}.}
\label{Fig6}
\end{figure}

 \noindent We find the correlation plots of baryon asymmetry of universe $Y$ with Dark matter mass $M_1$, effective Majorana mass  $|M_{ee}|$ and relic density of DM as shown in Fig. \ref{Fig1} to Fig. \ref{Fig6}. We plot baryon asymmetry of universe with relic density of DM to check the simultaneity of DM and BAU by considering 4 TeV as upper bound on DM mass. The horizontal lines in the given correlation plots corresponding to the observed value of the baryon asymmetry of the universe $Y=(6.04\pm 0.08)\times10^{-10}$ and the vertical line in Fig. \ref{Fig3} and Fig. \ref{Fig6} is the observed value of relic density of DM $\Omega h^2=0.120\pm 0.001$ \cite{Planck:2018vyg}. Fig. \ref{Fig1}  shows the correlation between baryon asymmetry of universe $Y$ and dark matter mass $M_1$ for the hybrid texture $G_1$. The plot shows that the texture $G_1$ produce successful baryogenesis through leptogenesis for the allowed range of DM mass i.e. $1 TeV$ to $2.27 TeV$. Similarly Fig. \ref{Fig4} shows the correlation plots between $Y$ and $M_1$ for the hybrid textures $G_{2,..5}$ and we find that these textures  produce successful leptogenesis as they satisfy the baryon asymmetry seen by Planck experiment.   Fig. \ref{Fig2} shows the correlation between baryon asymmetry of universe $Y$ with effective Majorana mass $|M_{ee}|$ for texture $G_1$ which successfully produce correct baryon asymmetry for effective Majorana mass $\geq 0.052$ $eV$.  Fig. \ref{Fig5} shows the correlation plots for baryon asymmetry of universe $Y$ verses effective Majorana mass $|M_{ee}|$ for textures $G_{2...5}$. It is clear from the Fig. \ref{Fig5} that these textures produce correct baryon asymmetry of universe $Y=(6.04\pm 0.08)\times10^{-10}$ \cite{Planck:2018vyg} and  gives a lower bounds on the effective Majorana mass $|M_{ee}|$ as shown in Table \ref{tab4}.


\section{Conclusions}
We investigate the leptogenesis, dark matter and neutrinoless double beta decay in the hybrid textures of neutrino mass matrix at one loop level. There are five hybrid textures which simultaneously accounts for the DM and neutrinoless double beta decay decay. We extend this idea to the simultaneous study of leptogenesis DM and neutrinoless double beta decay. We use the neutrino oscillation data and find the Yukawa couplings and use them to find the relic density of DM ($\Omega h^2$) and baryon asymmetry of universe. The correlation plots for these quantities shown in Fig. \ref{Fig1}, Fig. \ref{Fig2}, Fig. \ref{Fig4} and Fig. \ref{Fig5}. Fig. \ref{Fig1} and Fig. \ref{Fig4} show the predicted parameter space between baryon asymmetry of universe $Y$ and DM mass $M_1$ for the textures $G_{1...5}$. Therefore, from these figures we find that all these textures $G_{1...5}$ produce correct baryon asymmetry $Y=(6.04\pm 0.08)\times10^{-10}$\cite{Planck:2018vyg}. Fig. \ref{Fig2} and Fig. \ref{Fig5} show the predicted parameter space for the baryon asymmetry of universe $Y$ and effective Majorana mass $|M_{ee}|$ for the textures $G_{1...5}$. From these figures we find that the textures $G_{1...5}$ produce correct baryon asymmetry  and give lower bound to the effective Majorana mass $|M_{ee}|$. The most interesting plots are the correlation plots shown in Fig. \ref{Fig3} and Fig. \ref{Fig6} between baryon asymmetry of universe ($Y$) and relic density of dark matter ($R$). These plots are of more interest because confirm the simultaneity of the baryon asymmetry of universe $Y$ and dark matter $M_1$ with the determination of $R$. The textures $G_2$ and $G_3$ have been found to either satisfy the relic density bound $\Omega h^2 = 0.120 \pm 0.001$ or the bound on baryon asymmetry of the universe $Y=(6.04 \pm 0.08) \times 10^{-10}$  shown in the first row of Fig. \ref{Fig6} as there is no point satisfying both bounds, i.e. on or around the intersection of both bounds marked by small circles. Therefore, these textures (i.e. $G_2$ and $G_3$) do not show the simultaneity of the BAU and DM hence, discarded. Therefore, it has been found that out of five textures  $G_{1...5}$ only textures $G_1$, $G_4$ and $G_5$ simultaneously account for the leptogenesis, dark matter and neutrinoless double beta decay. Another interesting fact to note that is interesting to note Fig. \ref{Fig2} and Fig. \ref{Fig5} give lower bound to the $|M_{ee}|$ shown in Table \ref{tab4}. More evidence for or against the theories can be provided by the observation of $|M_{ee}|$ in past, present, and future neutrinoless double beta decay experiments \cite{Barabash:2011row,KamLAND-Zen:2016pfg,NEXT:2013wsz,NEXT:2009vsd,Licciardi:2017oqg}. These tests can explore smaller values of $|M_{ee}|$ with previously unheard-of sensitivity. The experiments e.g. SuperNEMO, KamLAND-Zen, NEXT, and nEXO (5 year) have sensitivity reaches of 0.05 eV, 0.045 eV, 0.03 eV, and 0.015 eV, respectively \cite{Barabash:2011row,KamLAND-Zen:2016pfg,NEXT:2013wsz,NEXT:2009vsd,Licciardi:2017oqg}.
    \begin{table}[H]
  \centering
\begin{tabular}{|c|c|c|c|c|}
 \hline
 Texture & $M_1 (TeV)$ & $|M_{ee}| (eV)$ & Leptogenesis\\
 \hline
$G_{1}$& $\leq 2.27$ & $\geq 0.052$& \cmark \\ 
\hline
%$G_{2}$ &  \multicolumn{2}{|l|}{disallowed by the observed value of relic density of DM($\Omega h^2$)} \tabularnewline  
% \hline
$G_{2}$&$\leq  5.31$ &  \multicolumn{2}{|l|}{disallowed by the observed value of relic density of DM($\Omega h^2$)and BAU ($Y$)} \tabularnewline
 \hline
$G_{3}$ & $\leq 3.93$  &   \multicolumn{2}{|l|}{disallowed by the observed value of relic density of DM($\Omega h^2$) and BAU ($Y$)} \tabularnewline
 \hline
$G_{4}$ & $\leq  2.77$ & $\geq 0.021$ &\cmark   \\
 \hline
$G_{5}$ & $\leq 4.10$&  $\geq 0.034$ &\cmark \\
 \hline
 \end{tabular}
 \caption{Ranges considered for the DM mass and bounds on effective Majorana mass  $|M_{ee}|$ for allowed hybrid textures.}
 \label{tab4}
 \end{table}

\noindent \textbf{\Large{Acknowledgments}}

 \noindent Ankush acknowledges the financial support provided by the University Grants Commission, Government of India vide registration number 201819-NFO-2018-19-OBC-HIM-75542. R. Verma acknowledges the financial support provided by the Central University of
Himachal Pradesh. B. C. Chauhan is thankful to the Inter University Centre for
Astronomy and Astrophysics (IUCAA) for providing necessary facilities during the
completion of this work.

\end{document}